\documentstyle[preprint]{ptptex}

\title{Exponentially small oscillation of 2-dimensional stable and
  unstable manifolds in 4-dimensional symplectic mappings}
\author{Yoshihiro Hirata\thanks{E-mail address:
    yhirata@allegro.phys.nagoya-u.ac.jp}, Kazuhiro Nozaki and Tetsuro
  Konishi}
\inst{Department of Physics, Faculty of Science, Nagoya University,
  Furo-cho, Chikusa-ku, Nagoya, 464-8602, Japan}
\abst{Homoclinic bifurcation of 4-dimensional symplectic mappings is
asymptotically studied.
We construct the 2-dimensional stable and unstable manifolds near the
sub-manifolds which experience exponentially small splitting, and
successfully obtain exponentially small oscillating terms in the
2-dimensional manifolds.
}
\preprintnumber{DPNU--99--01\\January 1999}

\def\a{\alpha}

\def\e{\epsilon}

\def\D{\Delta}

\def\DD{\D^2}

\def\Pj{\Phi_j}

\def\tj{t_j}
\def\ta{t_1}
\def\tb{t_2}

\def\qa{q_1}
\def\qb{q_2}
\def\qj{q_j}
\def\qk{q_k}
\def\pj{p_j}

\def\qjp{q_j^{\scriptscriptstyle \prime}}

\def\pjp{p_j^{\scriptscriptstyle \prime}}

\def\qjt{\qj (t)}
\def\qkt{\qk (t)}
\def\qt{q (t)}

\def\qjtt{\frac{d^2\qj}{dt^2}}

\def\qjte{\qj (t,\e )}
\def\qjttj{\qj (t,\tj )}

\def\st{s (t)}
\def\sdt{\dot{s}(t)}
\def\sj{s_j}
\def\sjt{s_j (t)}
\def\sjttj{s_j (t,\tj )}
\def\sjtzero{s_j (t,0)}
\def\sjdt{\dot{s}_j(t)}
\def\sjdttj{\dot{s}_j(t,\tj )}
\def\skt{s_k(t)}

\def\gt{g(t)}
\def\gjt{g_j(t)}
\def\gjttj{g_j(t,\tj )}

\def\qtjt{\tilde{q}_j(t)}
\def\qtj{\tilde{q}_j}

\def\qtt{\tilde{q}(t)}

\def\qtpmt{\tilde{q}^\pm (t)}
\def\qtjpmttj{\tilde{q}_j^\pm (t,\tj )}
\def\qtjpmtzero{\tilde{q}_j^\pm (t,0)}

\def\qjpmte{q_j^\pm (t;\e )}
\def\qjpmttj{q_j^\pm (t,\tj )}

\def\tcjpm{\tilde{c}_j^\pm}

\def\dqtt{\dot{\tilde{q}}(t)}

\def\tauj{\tau_j}

\def\Pjztj{\Phi_j(z,\tauj )}

\def\Pj{\Phi_j}
\def\Pk{\Phi_k}

\def\Xz{X(z)}
\def\tXz{\tilde{X}(z)}

\def\Xjpz{X^\prime_j(z)}
\def\Xkpz{X^\prime_k(z)}
\def\Xapz{X^\prime_1(z)}
\def\Xbpz{X^\prime_2(z)}

\def\qhjpmttj{\hat{q}^\pm_j(t,\tj )}
\def\qhzerot{\hat{q}^\pm_0(t)}
\def\qhpmonet{\hat{q}^\pm_1(t)}

\def\Mte{M(t,\e )}
\def\Bte{B(t,\e )}

\def\Mtte{\tilde{M}(t,\e )}
\def\Btate{\tilde{B}_1(t,\e )}
\def\Btbte{\tilde{B}_2(t,\e )}

\def\Ste{S(t,\e )}

\def\Z{{\bf Z}}

\def\del{\partial}

\def\D{\Delta}
\def\DD{\D^2}

\def\sech#1{{\rm sech}(#1)}

\def\sinh#1{{\rm sinh}(#1)}

\def\cosh#1{{\rm cosh}(#1)}

\def\O#1{O(#1)}

\def\exp#1{e^{#1}}

\makeatletter
\def\@begintheorem#1#2{\rm \trivlist
     \item[\hskip \labelsep{\bf #1\ #2}] \item[]}
\def\@opargbegintheorem#1#2#3{\rm \trivlist
     \item[\hskip \labelsep{\bf #1\ #2\ (#3)}] \item[]}
\makeatother

\def\ds{\displaystyle}

\begin{document}

\maketitle

Homoclinic (or heteroclinic) bifurcation plays an important role to
cause chaotic motion near hyperbolic fixed points.
Melnikov's method (or Melnikov's integral), which is based on the
regular perturbation method, is a powerful tool to detect transversal
homoclinic points.\cite{GH} \ 
On the other hand, it is known that in the rapidly forced systems and
the standard map, and so on, splitting of separatrices is
exponentially small.
Consequently it is impossible to capture exponentially small
oscillation straightforwardly with Melnikov's method.

In the last decade the difficulty has been overcome by using the
method called {\it asymptotic expansions beyond all orders}.
This method enables one to capture exponentially small oscillation of
the stable and unstable manifolds in 2-dimensional symplectic mappings
which are perturbed from linear mappings,\cite{LST89,GLS94,HM93} \ and
to construct functional approximations of the
manifolds.\cite{TTJ94,TTJ98,NH96}

For 4-dimensional symplectic mappings, Gelfreigh and Shatarov treated
a coupled standard mapping and computed the crossing angle between the 
2-dimensional stable and unstable manifolds.\cite{GS95} \ 
We obtained the functional approximations of particular sub-manifolds
of the stable and unstable manifolds in a 4-dimensional double-well
symplectic mapping with a weak coupling.\cite{HK} \ 
In this letter, we construct the functional approximations of
2-dimensional stable and unstable manifolds in 4-dimensional
double-well symplectic mappings with more general weak couplings.

We start with the 4-dimensional symplectic map $(\qj,\pj ) \mapsto
(\qjp,\pjp )$
\begin{eqnarray}
  \label{eqn:map} \left\{
  \begin{array}{rcl}
    \pjp & = & \ds \pj + \e ( \qj - 2\pj^3 ) - \e^3
    \frac{\del\tilde{J}}{\del\qj} \\
    \qjp & = & \qj + \e\pjp
  \end{array} \right.
\end{eqnarray}
where $j, k \in \{ 1, 2 \}, j \ne k$, $0 < \e \ll 1$ is a perturbation
parameter.
To simplify computation we take the coupling potential $\tilde{J}$ as
$\ds \tilde{J} = \frac{a}{2}\qa^2\qb^2 +
\frac{b}{3}\qa^3\qb^3$.\footnote{In Ref.~\citen{HK} we show that if
  $\tilde{J}$ is a polynomial of $\qj$'s of degree six, it is possible
  to capture exponentially small oscillation.
  Hence this coupling term is considerably general.}
Note that the map (\ref{eqn:map}) possesses the substitutional
invariance between $j$ and $k$.

The map (\ref{eqn:map}) possesses a hyperbolic $\times$ hyperbolic
fixed point at the origin $(0, 0, 0, 0 )$ and there exist
2-dimensional stable and unstable manifolds near the origin.
In Ref.~\citen{HK}, we applied the method to the particular
1-dimensional stable and unstable sub-manifolds defined as $q_1 = q_2$ 
and successfully obtained the functional approximations of the
sub-manifolds when $b = 0$.
In this letter we study the neighborhood of the sub-manifolds as a
perturbation problem.

One can rewrite the map (\ref{eqn:map}) in the second order difference
equation
\begin{eqnarray}
  \label{eqn:DifEqu}
  \D_\e^2 \qjt & = & \qjt - 2\qjt^3 - \e^2 ( a\qjt\qkt^2 +
  b\qjt^2\qkt^3 ),
\end{eqnarray}
where $\D_\e^2 \qt \equiv \{ q( t + \e ) - 2q(t) + q( t - \e
)\}/\e^2$.
The difference equation (\ref{eqn:DifEqu}) is equivalent to the
infinite order differential equation
\begin{eqnarray}
  \label{eqn:Outer}
  \qjtt & = & \qj - 2\qj^3 - \e^2 ( a\qj\qk^2 + b\qj^2\qk^3 ) - 2
  \sum_{l=2}^{\infty} \frac{\e^{2l-2}}{(2l)!} \frac{d^{2l}\qj}{dt^{2l}},
\end{eqnarray}
which is called {\it the outer equation}.
The boundary condition with which we solve the equation
(\ref{eqn:Outer}) is $\qjpmte \to +0\,(t \to \pm \infty)$,
where the superscript $\pm$ stands for the stable and unstable
manifolds, respectively.\footnote{The same notations will be used
  without further comment.}

We first construct the solution to the ODE (\ref{eqn:Outer}) by using
the regular perturbation method with respect to $\e^2$, i.e.,
expanding $\qjte = \sjt + \e^2 \qtjt + \O{\e^4}$.
Substituting the expansion into the ODE (\ref{eqn:Outer}), one obtains
the unperturbed equation as
\begin{eqnarray}
  \label{eqn:UPeq}
  \frac{d^2\sj}{dt^2} & = & \sj - 2\sj^3, \nonumber
\end{eqnarray}
and the perturbed equation as
\begin{eqnarray}
  \label{eqn:outerP}
  \frac{d^2\qtj}{dt^2} & = & ( 1 - 6 \sjt^2 )\qtj + F_j(t),
\end{eqnarray}
where $F_j(t)$ is a polynomial of $\sjt, \skt$ and their
derivatives.
The boundary condition is rewritten as $\sjt, \qtjt \to +0\ (t
\to \pm \infty)$.

The unperturbed solution is obtained as $\sjttj = \sech{t+\tj}$, where
$\tj$'s are integration constants.
Because the system is autonomous, it has translational invariance for
$t$.
Hence the essential parameter is the difference between $\ta$ and
$\tb$.
We put $\tj = (-1)^{j} \a$ where $\a$ is an arbitrary real constant.
To simplify notations we put $\st \equiv \sjtzero = \sech{t}$.
Taking the integration constant $\a$, which was vanished in
Ref.~\citen{HK}, as another perturbation parameter, we successfully
obtain exponentially small oscillating terms in the neighborhood of
the sub-manifolds.

The fundamental system of the solution to the homogeneous equation of
the ODE (\ref{eqn:outerP}) is given as $\sjdttj$ and $\gjttj = g(t+\tj
)$, where $\ds \gt = \sdt \int \frac{dt}{\sdt^2} = \frac{3}{2}\sech{t}
- \frac{1}{2}\cosh{t} - \frac{3}{2}t\, \sinh{t}\sech{t}^2$.
Note that $\sdt$ and $\gt$ are odd and even functions of $t$
respectively, and Wronskian is equal to $1$.
The general stable and unstable solutions to the inhomogeneous
equation (\ref{eqn:outerP}) is obtained as
\begin{eqnarray}
  \label{eqn:outerGS}
  \qtjpmttj & = & c_j^\pm \sjdt - \sjdt \int_{0}^{t}{F_j(t)
    \gjt dt} + \gjt \int_{\pm\infty}^{t}{F_j(t) \sjdt dt},
\end{eqnarray}
respectively, where $c_j^\pm$'s are integration constants.

When $\a = 0$, which is equivalent to $\tj = 0, j=1,2$, the solution
stands for the 1-dimensional sub-manifolds of the 2-dimensional stable 
and unstable manifolds, and the stable and unstable sub-manifolds
coincide, i.e., $q_j^+(t) = q_j^-(-t)$.\cite{HK} \ 
One can therefore restrict the solution to even fuctions of $t$.
Putting $\qtjpmtzero = \qtt$, one obtains
\begin{eqnarray}
  \label{eqn:outerPss}
  \qtt & = & \frac{b+2}{6} \sech{t}^3 - \frac{6a+8b+7}{24} \sech{t}
  + \frac{t}{24} \sinh{t}\sech{t}^2. \nonumber
\end{eqnarray}

Next we construct solutions when $\tj \ne 0$.
Let us suppose $| \tj | \ll 1$, corresponding to the neighborhood of
the sub-maifolds.
We expand $\qtjpmttj$ in the power of $\tj$ around $\qtt$ as
\begin{eqnarray}
  \qtjpmttj & = & \qtt + \sum_{j=1}^{2} \tj \left. \frac{\del
      \qtjpmttj}{\del \tj}\right|_{\tj = 0} + \O{\tj^2} \nonumber \\
  \label{eqn:2dimExpand}
  & = & \qtt + \tj \left( \frac{\ta}{\tj} \frac{\del
      \qtjpmtzero}{\del \ta} + \frac{\tb}{\tj} \frac{\del
      \qtjpmtzero}{\del \tb} \right) + \O{\tj^2},
\end{eqnarray}
and $c_j^\pm = c^0 + \tj \tcjpm + \O{\tj^2}$.
The condition $\qtjpmttj \to \qtt\ (\tj \to 0)$ implies $c^0 = 0$.
The terms in the bracket of the right hand side of the equation
(\ref{eqn:2dimExpand}) are independent of $j$ because of the
substitutional invariance between $j$ and $k$ of the map
(\ref{eqn:map}).
Hence we put them $\qtpmt$, i.e., $\qtjpmttj = \qtt + \tj \qtpmt$.
A straightforward computation gives
\begin{eqnarray}
  \label{eqn:outersolP}
  \qtpmt & = & \tcjpm \sdt \mp \frac{56a+48b}{105} \gt + \dqtt
  \nonumber \\
  & & + \frac{3}{7} b\, \sinh{t}\sech{t}^4 - \frac{28a+24b}{35}
  \sinh{t} \sech{t}^2 \log{(\cosh{t})} \nonumber \\
  & & + \frac{8a-6b}{30} \sinh{t} \sech{t}^2 - \frac{28a+24b}{105}
  \sinh{t}. \nonumber
\end{eqnarray}
Thus the solution to the outer equation (\ref{eqn:Outer}) is given as
\begin{eqnarray}
  \label{eqn:outerResult}
  \qjpmttj & = & \st + \e^2 \qtt + \tj \{ \sdt + \e^2 \qtpmt \}
\end{eqnarray}
up to $\e^4$ and $\tj^2$.

The outer solution (\ref{eqn:outerResult}) possesses singular points
at $t = \pi i/2 + n\pi i,\ n\in \Z$.
One must therefore analyze the neighborhood of the points as the inner
problem.
Introducing the new variables as $t - \pi i/2 = \e z$, $\tauj = \tj
/\e$ and $\Pjztj = \e\qjttj$, the difference equation
(\ref{eqn:DifEqu}) is rewritten as $\DD \Pj = -2 \Pj^3 - b \Pj^2 \Pk^3
+ \e^2 ( \Pj - a \Pj \Pk^2)$, where $\DD \Pj \equiv \Phi_j(z+1,\tauj )
-2\Pjztj + \Phi_j(z-1,\tauj )$.
In this letter we restrict our attention to the leading term of
$\Pjztj$, hence truncating the terms with respect to $\e^2$, we
obtain {\it the inner equation} as
\begin{eqnarray}
  \label{eqn:Inner}
  \DD \Pj & = & -2 \Pj^3 - b \Pj^2 \Pk^3.
\end{eqnarray}
Note that when $b \ne 0$, the second term, which is generated from the
coupling term, contribute the inner equation.

When $\a = 0$, putting $\Xz = \Pj (z,0)$, the inner equation
corresponing to $\st + \e^2 \qtt$ is obtained as
\begin{eqnarray}
  \label{eqn:inner1}
  \DD \Xz & = & -2\Xz^3 - b\Xz^5,
\end{eqnarray}
which is called the first inner equation.
The matching condition is $\Xz \to \e (\st + \e^2\qtt )$ when $|z| \to 
\infty$.
On the other hand, when $\a \ne 0$, expanding $\Pj (t,\tauj ) = \Xz +
\tauj \Xjpz$, we obtain
\begin{eqnarray}
  \label{eqn:PreInner}
  \DD \Xjpz & = & - 6 \Xz^2 \Xjpz - b \Xz^4 \{ 2 \Xjpz - 3 \Xkpz \}.
\end{eqnarray}
Summing up both sides of the equations (\ref{eqn:PreInner}), one
obtains the second inner equation
\begin{eqnarray}
  \label{eqn:inner2}
  \DD \tXz & = & -6\Xz^2\tXz + b\Xz^4\tXz,
\end{eqnarray}
where $\tXz = \Xapz + \Xbpz$.\footnote{The subtraction of both sides
  of the equations (\ref{eqn:PreInner}) gives the linearized equation
  of the equation (\ref{eqn:inner1}).
  In this computation the linearized equation has no
  information.\cite{HNK99}}
The matching condition is $\tXz \to \e^2(\sdt + \e^2\qtpmt)$ when $|z|
\to \infty$.

The inner equations (\ref{eqn:inner1}) and (\ref{eqn:inner2}) possess
formal solutions when $|z|$ tends to $\infty$.
Although the formal solutions do not converge, the formal Borel
transformations of these formal solutions converge, i.e., these formal 
solutions are Borel summable.
By using Borel transformation, Stokes phenomena, which are abrupt jump
of asymptotic expansions, around $t = \pi i/2$, can be captured, and
the analytic continuation from the left half of $t$-plane to the right
one can be performed.
The details of the computation will be reported elsewhere.\cite{HNK99}

Finally the stable and unstable solutions which are valid both the
left and right half plane are obtained as
\begin{eqnarray}
  \label{eqn:res}
  \qhjpmttj = \qhzerot + \tj\qhpmonet,
\end{eqnarray}
where
\begin{eqnarray}
  \qhzerot & = & \st + \e^2 \qtt + \Ste\Mte\Bte \nonumber \\
  \label{eqn:qzero}
  \qhpmonet & = & \sdt + \e^2 \qtpmt + \Ste\{ \Mte \Btate + \Mtte
  \Btbte \}\\
  \Mte & = & - \frac{5}{2}\frac{c}{\e^4} \exp{-\frac{\pi^2}{\e}}
  \nonumber \\
  \Mtte & = & - \frac{5}{2}\frac{\tilde{c}}{\e^5}
  \exp{-\frac{\pi^2}{\e}} \nonumber \\
  \Bte & = & \frac{3}{4} \pi \sdt \cos{\frac{2\pi t}{\e}} + \gt
  \sin{\frac{2\pi t}{\e}} \nonumber \\
  \Btate & = & \frac{3}{4} \pi \ddot{s}(t) \cos{\frac{2\pi t}{\e}} +
  \dot{g}(t) \sin{\frac{2\pi t}{\e}} \nonumber \\
  \Btbte & = & \frac{3}{4} \pi \sdt \sin{\frac{2\pi t}{\e}} - \gt
  \cos{\frac{2\pi t}{\e}} \nonumber \\
  S(t,\e ) & = & \left\{
    \begin{array}{rl}
      0 & ( t < 0 )\\
      1 & ( t = 0 )\\
      2 & ( t > 0 )
    \end{array} \right.{({\rm for\ small}\ \e )}. \nonumber
\end{eqnarray}
The Stokes constants $c$ and $\tilde{c}$, which depend on $b$, are
numerically obtained as $c = -763$ and $\tilde{c} = 1146$ when $b =
1.0$.
For other values of $b$ the Stokes constants also converge.\cite{HNK99}

By using the approximate solution (\ref{eqn:res}), the splitting
angles between the stable and unstable manifolds can be
computed.\cite{HNK99} \ 
One can show that the angle along the sub-manifolds ($\ta = \tb$) is
exponentially small\cite{NH96} \ and the angle orthogonal to the
sub-manifolds are $\O{\e^2}$.

The expression (\ref{eqn:qzero}) implies that not only the
sub-manifolds but also the 2-dimensional stable and unstable manifolds
themselves possess exponentially small oscillating terms.
In the future work, we extend this computation to $2n$-dimensional
symplectic mappings with weak coupling terms.

\end{document}